\documentstyle[12pt,psfig]{article}

\newcommand{\appendixA}{\setcounter{equation}{0}
\def\theequation{\rm{A}.\arabic{equation}}\section*}

\catcode`\@=11
\def\marginnote#1{}

\def\ifmath#1{\relax\ifmmode #1\else $#1$\fi}

\def\bold#1{\setbox0=\hbox{$#1$}%
     \kern-.025em\copy0\kern-\wd0
     \kern.05em\copy0\kern-\wd0
     \kern-.025em\raise.0433em\box0 }

\def\GENITEM#1;#2{\par\vskip6pt \hangafter=0 \hangindent=#1
   \Textindent{$ #2$ }\ignorespaces}

\newcount\hour
\newcount\minute
\newtoks\amorpm
\hour=\time\divide\hour by60
\minute=\time{\multiply\hour by60 \global\advance\minute by-
\hour}
\edef\standardtime{{\ifnum\hour<12 \global\amorpm={am}%
    \else\global\amorpm={pm}\advance\hour by-12 \fi
    \ifnum\hour=0 \hour=12 \fi
    \number\hour:\ifnum\minute<100\fi\number\minute\the\amorpm}}
\edef\militarytime{\number\hour:\ifnum\minute<100\fi\number\minute}
\def\draftlabel#1{{\@bsphack\if@filesw {\let\thepage\relax
  \xdef\@gtempa{\write\@auxout{\string
    \newlabel{#1}{{\@currentlabel}{\thepage}}}}}\@gtempa
    \if@nobreak \ifvmode\nobreak\fi\fi\fi\@esphack}
     \gdef\@eqnlabel{#1}}
\def\@eqnlabel{}
\def\@vacuum{}
\def\draftmarginnote#1{\marginpar{\raggedright\scriptsize\tt#1}}
\def\draft{\oddsidemargin -.5truein
        \def\@oddfoot{\sl preliminary draft \hfil
        \rm\thepage\hfil\sl\today\quad\militarytime}
        \let\@evenfoot\@oddfoot \overfullrule 3pt
        \let\label=\draftlabel
        \let\marginnote=\draftmarginnote

\def\@eqnnum{(\theequation)\rlap{\kern\marginparsep\tt\@eqnlabel}%
\global\let\@eqnlabel\@vacuum}  }
\def\preprint{\twocolumn\sloppy\flushbottom\parindent 1em
        \leftmargini 2em\leftmarginv .5em\leftmarginvi .5em
        \oddsidemargin -.5in    \evensidemargin -.5in
        \let\@evenfoot\@oddfoot \overfullrule 3pt
        \let\label=\draftlabel
        \let\marginnote=\draftmarginnote

\def\@eqnnum{(\theequation)\rlap{\kern\marginparsep\tt\@eqnlabel}%
\global\let\@eqnlabel\@vacuum}  }
\def\preprint{\twocolumn\sloppy\flushbottom\parindent 1em
        \leftmargini 2em\leftmarginv .5em\leftmarginvi .5em
        \oddsidemargin -.5in    \evensidemargin -.5in
        \columnsep 15mm \footheight 0pt
        \textwidth 250mmin      \topmargin  -.4in
        \headheight 12pt \topskip .4in
        \textheight 175mm
        \footskip 0pt

\def\@oddhead{\thepage\hfil\addtocounter{page}{1}\thepage}
        \let\@evenhead\@oddhead \def\@oddfoot{} \def\@evenfoot{}
}
\def\titlepage{\@restonecolfalse\if@twocolumn\@restonecoltrue\onecolumn
     \else \newpage \fi \thispagestyle{empty}\c@page\z@
        \def\thefootnote{\fnsymbol{footnote}} }
\def\endtitlepage{\if@restonecol\twocolumn \else  \fi
        \def\thefootnote{\arabic{footnote}}
        \setcounter{footnote}{0}}  
\catcode`@=12
\relax

\def\lesssim{\stackrel{<}{\sim}}
\def\bigsim{\stackrel{>}{\sim}}

\def\be{\begin{equation}}
\def\ee{\end{equation}}
\def\br{\begin{eqnarray}}
\def\er{\end{eqnarray}}

\def\NPB#1#2#3{{\it Nucl.~Phys.} {\bf{B#1}} (19#2) #3}
\def\PLB#1#2#3{{\it Phys.~Lett.} {\bf{B#1}} (19#2) #3}
\def\PRD#1#2#3{{\it Phys.~Rev.} {\bf{D#1}} (19#2) #3}
\def\PRL#1#2#3{{\it Phys.~Rev.~Lett.} {\bf{#1}} (19#2) #3}
\def\ZPC#1#2#3{{\it Z.~Phys.} {\bf C#1} (19#2) #3}

\def\PRD#1#2#3{{\it Phys.~Rev.} {\bf{D#1}} (19#2) #3}
\def\PRL#1#2#3{{\it Phys.~Rev.~Lett.} {\bf{#1}} (19#2) #3}
\def\ZPC#1#2#3{{\it Z.~Phys.} {\bf C#1} (19#2) #3}

\begin{document}

\topmargin-2.5cm  

\begin{titlepage}
\begin{flushright}
SUSX-TH/97-010\\ 
IEM-FT-159/97\\
\end{flushright}
\vspace{.2in}
\begin{center}

{\large{\bf Weak Magnetic Dipole Moments in the
MSSM\footnote{Work partially supported by
the European Commission under the Human Capital and Mobility
programme, contract no.\ CHRX-CT94-0423.}}
}
\bigskip \\
{\large B.~de Carlos${}^a$\footnote{Work supported by PPARC.} and 
J.~M.~Moreno${}^b$\footnote{Work supported by CICYT of Spain,
contract no.\ AEN95-0195.}}\\ 
\vskip 0.2in
{\it 
${}^a$ Centre for Theoretical Physics, 
University of Sussex, \\ Falmer, Brighton BN1 9QH, UK. \\
Email: {\tt B.De-Carlos@sussex.ac.uk} \\}
\vskip 0.2in
{\it
${}^b$ 
Instituto de Estructura de la Materia, CSIC, \\
Serrano 123, 28006-Madrid, Spain. \\
Email: {\tt jmoreno@pinar1.csic.es}
}
\\ \vspace{.5in}
{\bf Abstract} \smallskip \end{center} \setcounter{page}{0}
We calculate the weak magnetic dipole moment of different 
fermions in the MSSM. In particular, we consider in detail the 
predictions for the WMDM of the $\tau$ lepton and bottom quark.
We compare the purely SUSY contributions with two Higss 
doublet models and SM predictions. For the $\tau$ lepton,
we show that chargino diagrams give the main SUSY contribution,
which for $\tan \beta  = 50$ can be one order of 
magnitude bigger than the SM prediction. For the $b$ quark, 
gluino diagrams provide the main SUSY contribution to its weak 
anomalous dipole moment, which is still dominated by gluon 
contributions. We also study how the universality assumption 
in the slepton sector induces correlations between the SUSY
contributions to the $\tau$ WMDM and to $(g-2)$ of the muon.

\end{titlepage}

\newpage
 
\section{Introduction}

One of the most promising candidates for physics beyond the so--called 
Standard Model (SM) is that of supersymmetry (SUSY) \cite{revs}. SUSY
has the highly attractive properties of giving a natural explanation
to the hierarchy problem of how it is possible to have a low energy 
theory containing light scalars (the Higgs) when the ultimate theory 
must include states with masses of order of the Planck mass. 

On the other hand, the spectrum of new particles predicted by SUSY seems 
to lie beyond the region explored by present colliders. The information 
coming from direct searches is nicely complemented by the one 
provided by precision measurements. The predicted values for these 
measurements are sensitive to supersymmetric, virtual, contributions. 
An example of this kind of observables is given by magnetic dipole 
moments (WMDM). 
In a renormalizable theory, they are generated by quantum corrections, 
and then virtual effects from new physics appear at the same 
level as SM weak contributions. Due to the chiral nature of these
observables, we expect that the induced corrections will be suppressed
by some power of $m_f/M$, where $m_f$ is the mass of the fermion 
and $M$ is the typical scale of new physics. In that case, heavy 
third generation fermions would be the preferred candidates 
to look for this kind of quantum corrections. 

In this paper we present a complete calculation of 
the WMDM of the $\tau$ lepton and bottom quark within 
the Minimal Supersymmetric Standard Model (MSSM) framework. 
We will compare the contributions from the three different 
sectors: the electroweak one, the two higgs doublet sector and 
the purely supersymmetric one involving charginos, neutralinos and 
sfermions. We will analyze the regions of the supersymmetric 
parameter space where these new contributions could be more relevant.
Finally, we will also study other observables such as $(g-2)_{\mu}$
that could get potentially large supersymmetric contributions
in the region where the SUSY corrections to the WMDM are enhanced.

\section{The $\tau$ lepton}

It is well known that Z peak data prove the existence of
quantum corrections to the vector and axial Z couplings 
to fermions at least up to 1-loop level.
But, as we comment before, in general these 1-loop contributions 
also induce a new effective magnetic moment type 
interaction\footnote{In this paper we will consider only
CP--conserving dipole moments. 
CP--violating electric dipole moments in SUSY theories 
are also very interesting, although their value is much more 
model dependent.} given by:
\be
{\cal L}_{W \! M \! D  \! M} = 
\frac{e}{2 m_f}\,
{\bar u}(p_-)
\left[ a_f(q^2)\, i \sigma^{\mu \nu} q_\nu  \right]
v(p_+) Z_ \mu ,
\label{eq:aF}
\ee
where $e$ is the positron electric charge
and $q$ stands for the momentum of the $Z$ boson. The conventions 
used for the different momenta are depicted in fig.~1.
\begin{figure}
\centerline{  
\psfig{figure=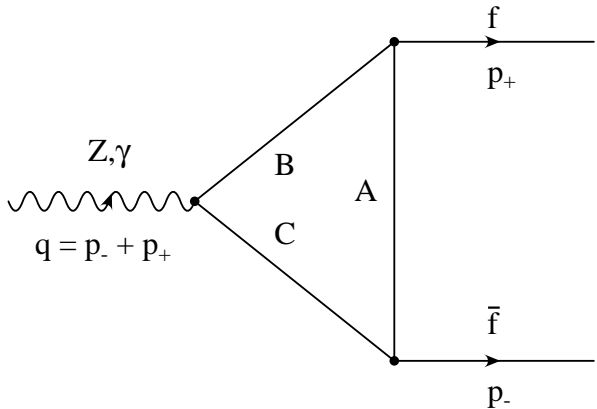,height=6cm,width=9cm,bbllx=0cm,bblly=7cm,bburx=21cm,bbury=21cm}
}
\caption{\footnotesize General diagram contributing to the WMDM of a fermion $f$.}
\end{figure}
This weak magnetic dipole moment, $a_f$, is a well defined,
gauge invariant quantity when the Z boson is on shell.

The different measurements on the Z peak, 
such as $\Gamma_{f \bar{f}}$, asymmetries, etc., 
provide {\em indirect} bounds on these WMDM \cite{Ano}.
Let us focus first on the $\tau$ lepton case.    
In particular, using the data presented at Moriond 1997 \cite{Mor97},
Rizzo gets\footnote{Notice that his definition of the
the anomalous magnetic moment, $\kappa^Z$, differs from
ours in a factor given by 
$ a^Z = {\displaystyle \frac{\kappa^Z}{2 s_Wc_W} } $.}
\cite{Ano} 

\be
 | {\rm Re}(a_\tau^Z) | <  0.0027 \;\; 95\% \; CL
\ee

In a general model, if the scale of the new interactions
generating the terms in eq.~(1) is sufficiently large,
then this operator will be contained in
some more general $SU(2) \times U(1)$ gauge invariant operators,
involving also $W$ and the higgs field \cite{EffLag}.
This correlation among the ``sizes'' of the WMDM operator and
their gauge companions allows to put extra bounds
on the magnetic and electric dipole moment values
(see, for example Escribano and Mass\'o in \cite{EffLag}.)
Notice, however, that in general the MSSM case cannot be described
in this framework. Therefore we will not consider here indirect $a_f$ 
bounds derived from these kind of relations.

On the other hand, L3 has recently presented the first {\em direct} 
limit on a WMDM, the one corresponding to this lepton.
They use correlated azimuthal asymmetries of the
$\tau^+\tau^-$ products, as proposed in \cite{BGV94}, and get 
\cite{sanchez}
\be
 | {\rm Re}(a_\tau^Z) | < 0.014  \;\;  95 \% \; CL
\ee

The maximum sensitivity expected on this WMDM from a complete analysis 
of LEP 1 data would be of the order of $10^{-4}$ \cite{BGTV95}. 
The SM prediction for $a_\tau$ is
\be
a_\tau^Z = -(2.10 + 0.61 i) \; 10^{-6} \;\;,
\ee
and was calculated by Bernab\'eu et {\em al.} \cite{BGTV95}. 
They showed that
$a_\tau^Z$ in the SM is dominated by $W-\nu$ diagrams.
The Higgs diagrams are suppressed by $\tau$ Yukawa coupling 
factors and their contribution is negligible for allowed $m_H$ values. 

Let us now briefly review the situation in two Higgs doublet 
models. As is well known, the corresponding $\tau-h^0$, $\tau-H^0$  
couplings are not necessarily small in these models, 
even if they are proportional to $m_\tau/M_Z$ .
In fact, they depend on the value of $\tan \beta$, the ratio  
between the vacuum expectation values of the two Higgs fields. It can 
be shown that for large values of $\tan \beta$
and scalar masses around $M_Z$, diagrams involving neutral Higgses 
can be of the same order of magnitude as the dominant SM contribution 
\cite{BCLS96}. The same considerations can be done for the new charged
Higgs contributions.
Therefore, we have to keep in mind that new diagrams in these models
could give contributions,  $a_\tau^Z|_{2HD}$, comparable to the SM result. 

Let us now analyze the situation in the MSSM. We can split the 
contributions to the WMDM as follows:
\be
a_\tau^Z = a_\tau^Z|_{EW} + a_\tau^Z|_{2HD} + a_\tau^Z|_{SUSY} \;\;,
\ee
where the new term $a_\tau^Z|_{SUSY}$ stands for the corrections 
given by chargino and neutralino diagrams. They are depicted in fig.~2.
As we said before, $a_\tau^Z|_{EW}$ is given in \cite{BGTV95}
and a detailed computation of $a_\tau^Z|_{2HD}$ can be found 
in \cite{BCLS96}. We will not repeat these results here, but
we want to stress that we have checked the analytical formulae 
and the numerical results presented in \cite{BCLS96}.
%
\begin{figure}
\centerline{  
\psfig{figure=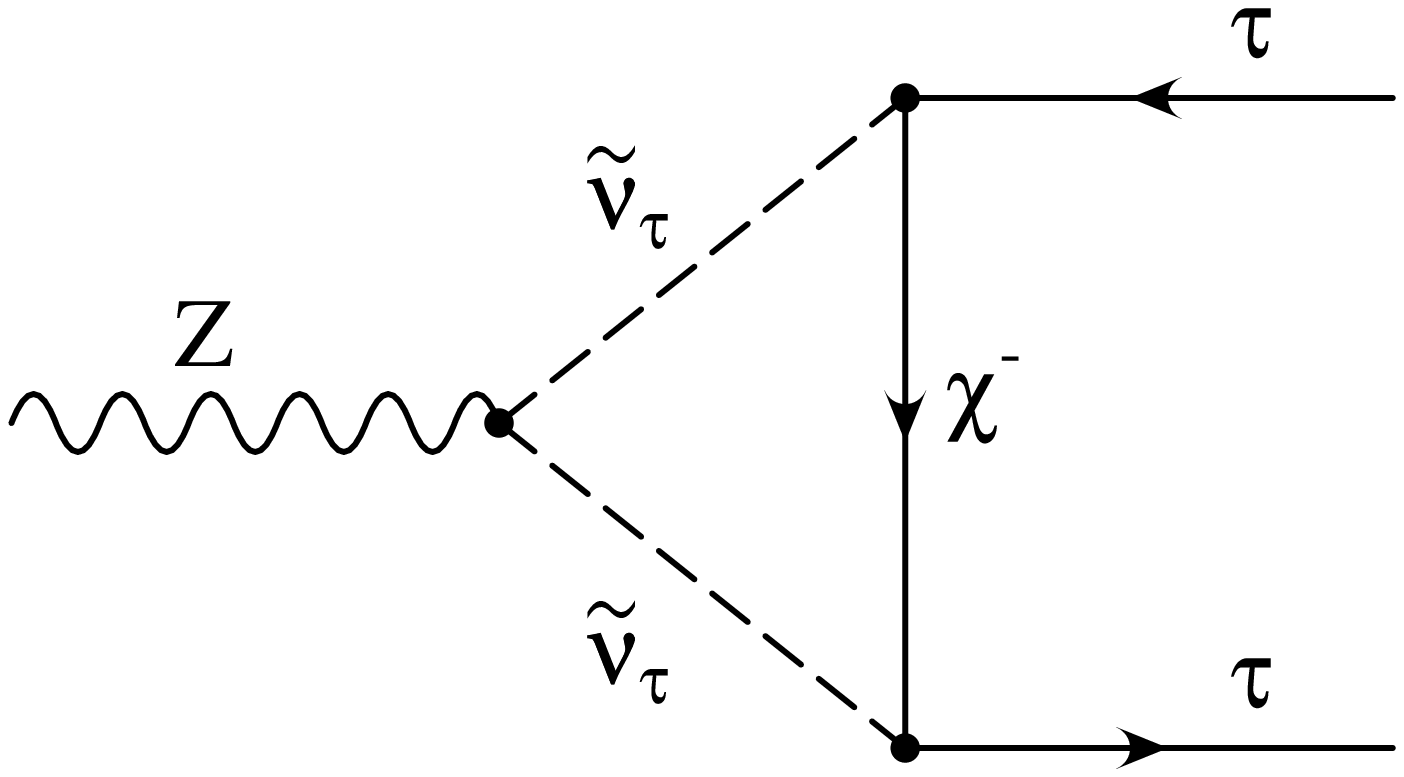,height=9cm,width=9cm,bbllx=0cm,bblly=7cm,bburx=21cm,bbury=21cm}\
\psfig{figure=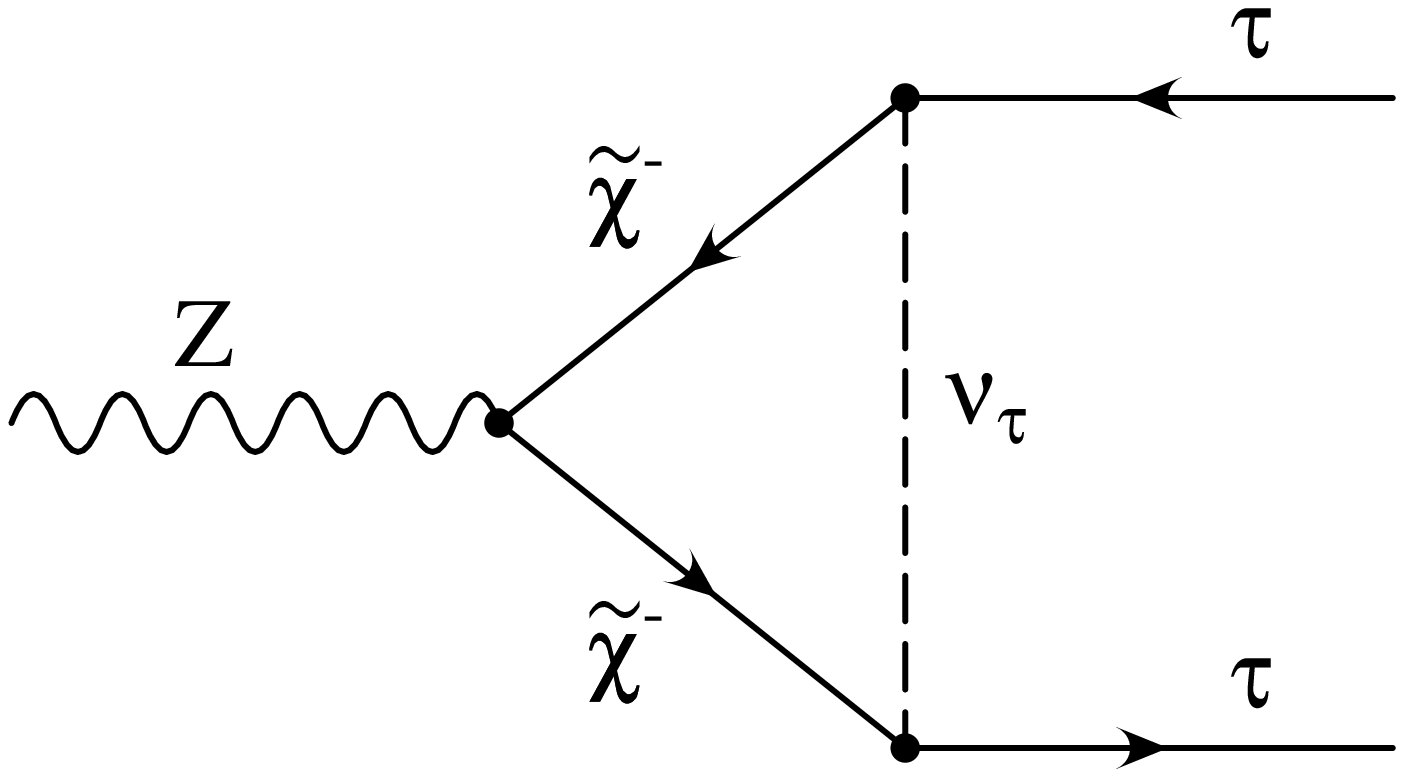,height=9cm,width=9cm,bbllx=0cm,bblly=7cm,bburx=21cm,bbury=21cm}\
}
\centerline{
\psfig{figure=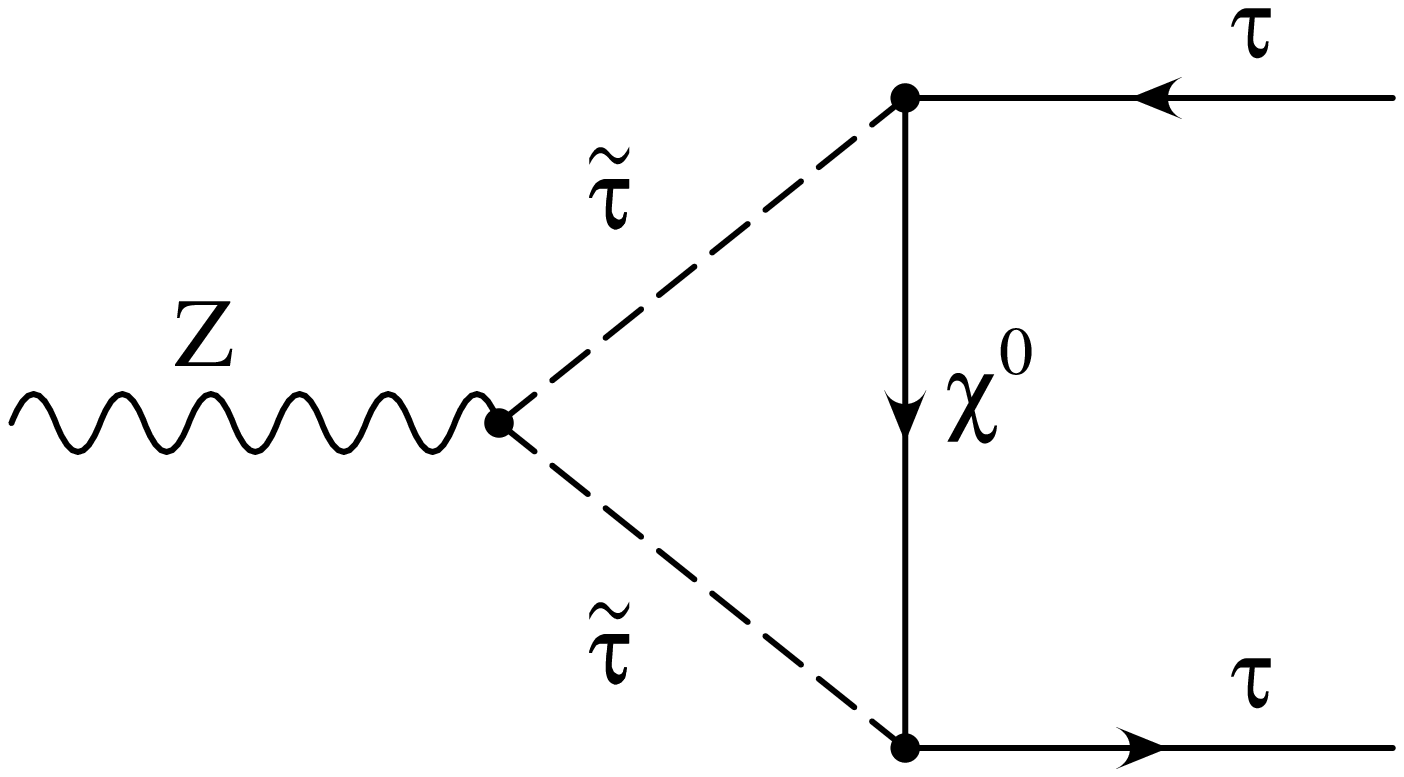,height=9cm,width=9cm,bbllx=0cm,bblly=7cm,bburx=21cm,bbury=21cm}\
\psfig{figure=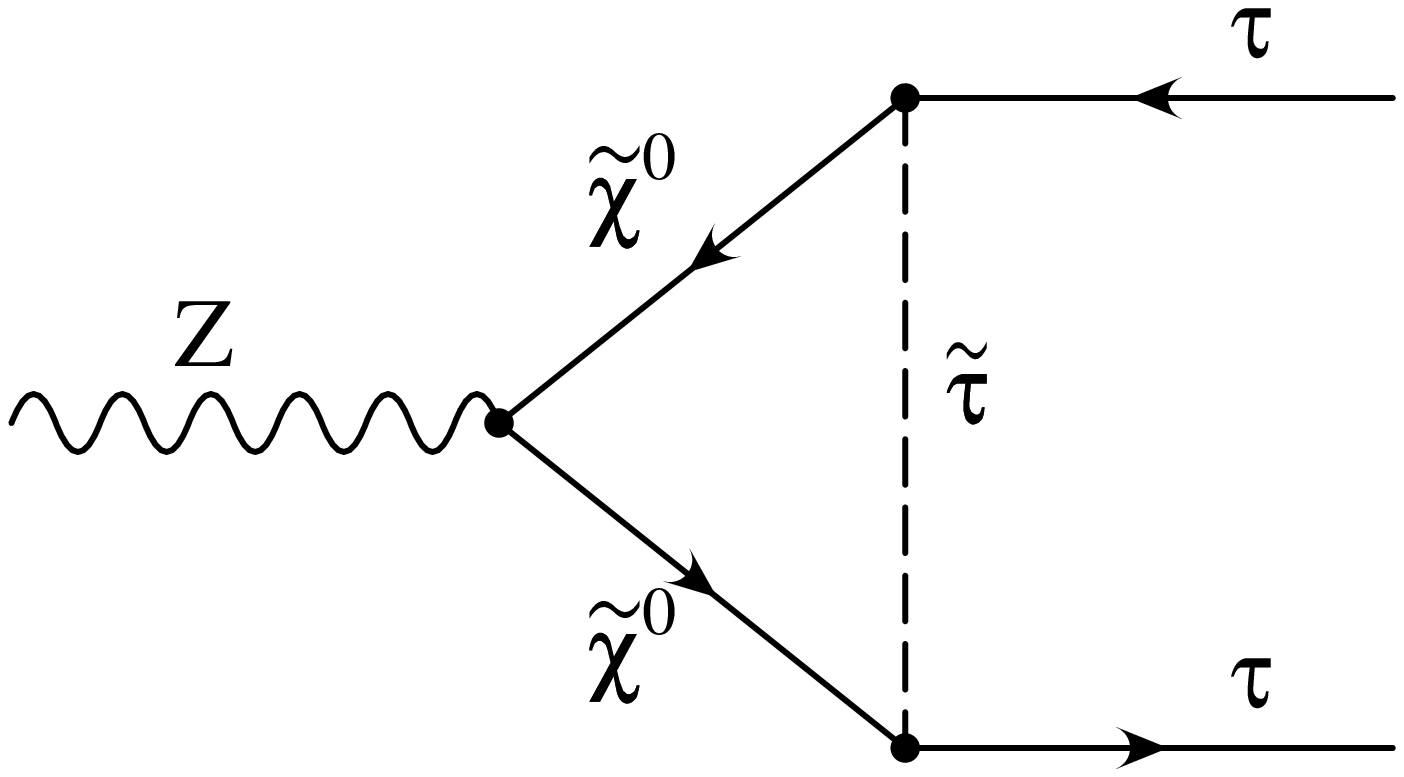,height=9cm,width=9cm,bbllx=0cm,bblly=7cm,bburx=21cm,bbury=21cm}
}
\caption{\footnotesize Relevant SUSY diagrams involving charginos and neutralinos
that contribute to the WMDM of the $\tau$ lepton.}
\end{figure}
%
We have collected in the appendix the expressions for the WMDM as 
functions of the SUSY masses and couplings.

\subsection{ SUSY results}

As we have seen, two sets 
of SUSY diagrams contribute to the WMDM: one with charginos and 
sneutrinos running in the loop and the other one with neutralinos and 
staus (see fig.~2). In general the chargino/sneutrino diagrams are 
going to dominate over the neutralino/stau ones, but not so 
overwhelmingly as to make the latter negligible, especially for low 
values of $\tan \beta$. 
Moreover both sets will sometimes contribute with opposite sign to
that of the SM+2HDM diagrams (which from now on will be denoted as 
``non-SUSY'' contributions), therefore potential cancellations can 
take place depending on the values of the different parameters. In
fact we have to distinguish between two different situations depending
on the sign of $\mu$. For the rest of this analysis we are going to 
assume values for the scalar soft masses, as well as for 
the trilinear couplings and $M_1$ (the soft bino mass), low enough 
(i.e. $\lesssim 300$ GeV) to give physical masses for the sparticles
(staus and sneutrinos in this particular case) and neutralinos close 
to their present experimental limits \cite{SUSYlimits}. Finally the
mass of the pseudoscalar, $m_A$, has been chosen so as to yield a
lightest neutral Higgs mass of approximately $80$ GeV.
Also, from now on we will give results for the {\em real} part of
the WMDM, given that the only relevant contribution to the imaginary
one would come from the lightest Higgs diagram and has already been
studied in \cite{BCLS96}. There could be some contribution coming from
those diagrams with neutralinos lighter than $M_Z/2$, but we have
checked that their effect on the total result is not important.

For $\mu>0$ and moderately small values of $\tan \beta$ (we have
chosen  $\tan \beta=2$ as a representative one) the chargino  
contribution has opposite sign to the non--SUSY one which, as said 
before, is going to lead to cancellations and therefore a total WMDM, 
in general, smaller that the one corresponding to the SM. On the other hand
the neutralino contribution has the same sign as that of the non--SUSY
one, but its absolute value is small enough to have no important 
effect on the final result.
This can be seen in the right half plane of fig.~3(a), where we plot
contour lines of equal WMDM (in units of $10^{-6}$) in the $\mu-M_2$
plane, which determines a lightest chargino mass of at most 250 GeV. 
For reasonably small values of the different sparticle masses we see 
that the total WMDM is never bigger than $-1.7.10^{-6}$, which
corresponds to $a_{non-SUSY} \sim -2.1.10^{-6}$, $a_{\chi^{\pm}}|_{min}
\sim 0.4.10^{-6}$ and $a_{\chi^0} \sim -1.10^{-8}$. This corresponds
to minimising the contribution of the charginos, which happens when
their masses increase. As we approach the contour line of 
$m_{\chi^{\pm}_1}=85$ GeV shown in the plot, the total WMDM decreases 
as $a_{\chi^{\pm}}$ increases reaching a maximum value of $9.10^{-6}$.
\begin{figure}
\centerline{
\psfig{figure=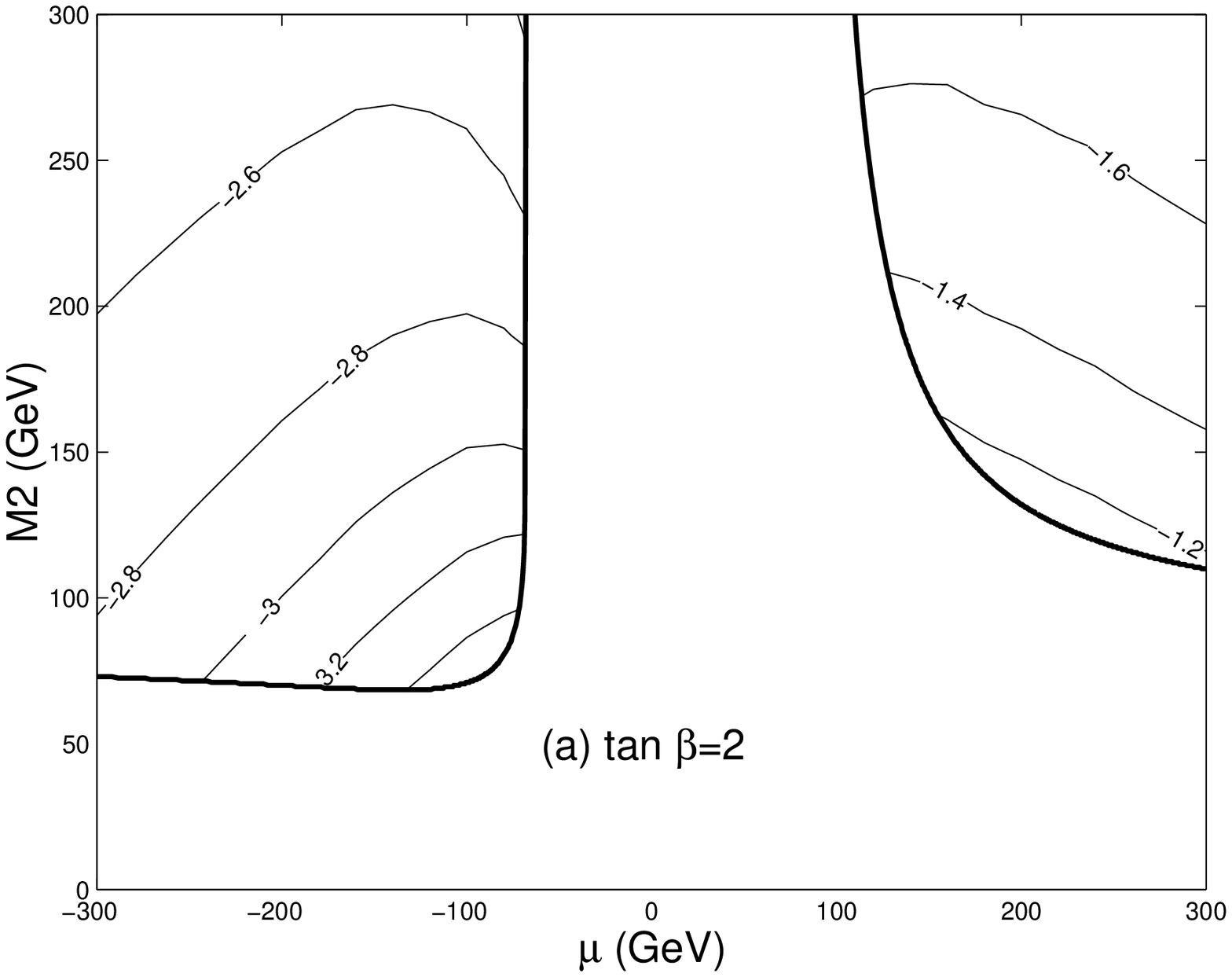,height=9cm,width=9cm,bbllx=0cm,bblly=7cm,bburx=21cm,bbury=21cm}\
\psfig{figure=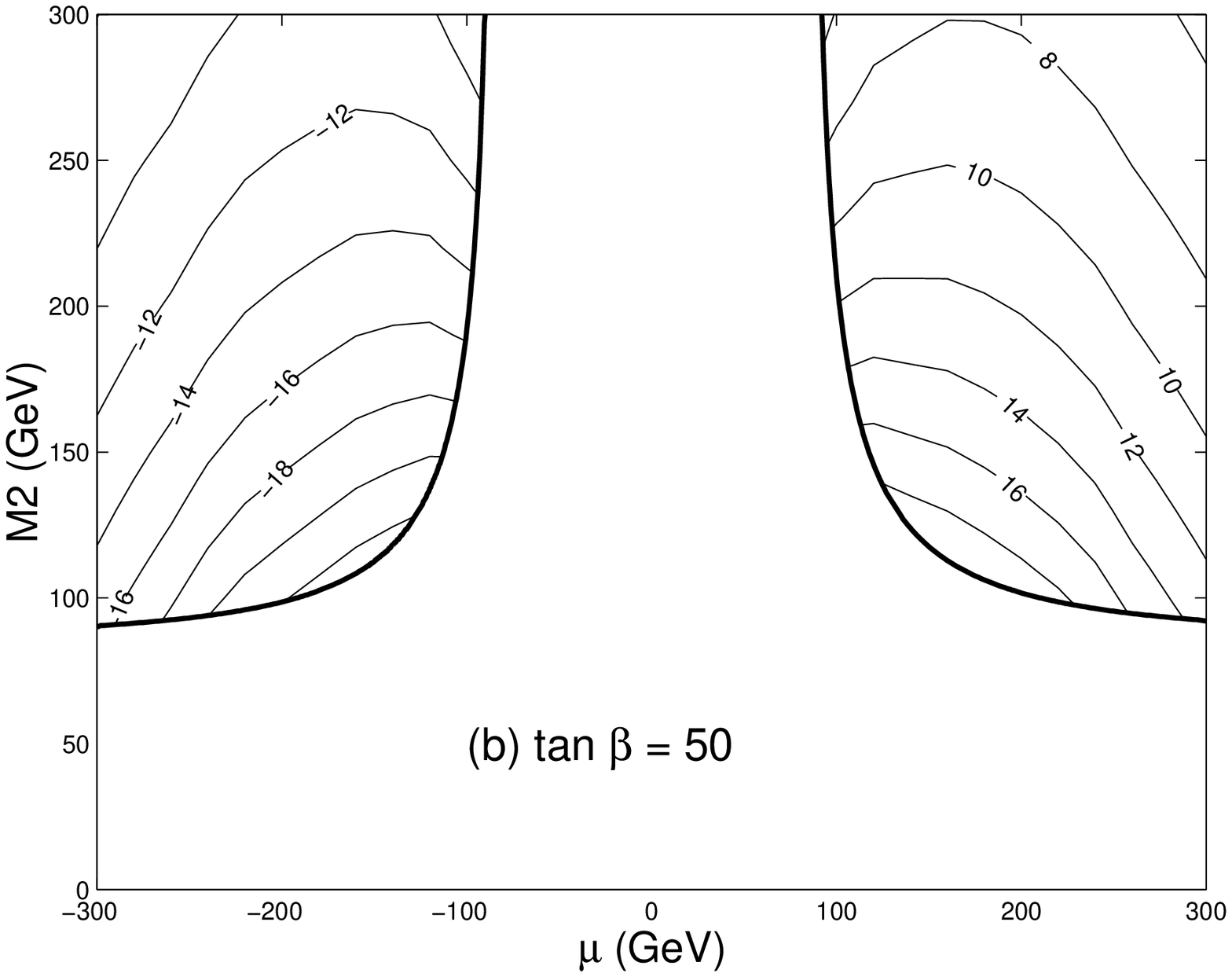,height=9cm,width=9cm,bbllx=0cm,bblly=7cm,bburx=21cm,bbury=21cm}
}
\caption{\footnotesize
Contour lines of constant WMDM (in units of $10^{-6}$) of the $\tau$ 
lepton in the $\mu$ vs $M_2$ plane for (a) $\tan \beta=2$; (b) $\tan 
\beta=50$. The thick line in both plots represents the contour 
$m_{\chi^{\pm}_1} =85$ GeV. 
}
\end{figure} 
Therefore the total result varies between $-1.1.10^{-6}$ and 
$-1.7.10^{-6}$.

Let's now turn to the most interesting case of $\mu<0$. Now the
chargino/sneutrino and neutralino/stau contributions have the same sign
as the non-SUSY one and, therefore, the total result is above the SM
prediction. This is shown in the left-hand side of fig.~3(a). As a
general feature, $a_{non-SUSY} \sim -2.1~.~10^{-6}$, whereas  
$a_{\chi^{\pm}}|_{max} \sim -1.3.10^{-6}$ and $a_{\chi^0} \sim
-1.10^{-7}$ giving a total WMDM of approximately $-3.5.10^{-6}$
along the line of $m_{\chi^{\pm}_1} = 85$ GeV. Similarly to the $\mu
>0$ case, as the chargino mass increases, its contribution, and
therefore in this case $a_{tot}$, decreases reaching a minimum value 
of $-2.10^{-6}$. 

In conclusion, for $\tan \beta=2$ we achieve some enhancement with 
respect to the SM for $\mu<0$, but not enough to be detected in the 
near future. Also, as a general feature throughout this analysis, 
increasing $\tan \beta$ will induce larger WMDM values.
This is due to the fact that
increasing $\tan \beta$ notably increases the chargino contribution;
therefore for $\mu>0$ this will result in a total WMDM of around 0
when $a_{\chi}^{\pm} \sim a_{non-SUSY}$ (which occurs for $\tan \beta
\sim 10$), and from then on, in a total WMDM increasingly dominated 
by the chargino diagrams. For $\mu<0$, as we said all the three 
contributions to the WMDM (namely non--SUSY, chargino and neutralino) 
have the same sign, and will always result in a total value bigger
than the SM one. This together with the fact that $a_{\chi^{\pm}}$ 
grows with increasing $\tan \beta$, will lead to values of the WMDM
well above the SM contribution from $\tan \beta \sim 10$ onwards.

Let's focus on  $\tan \beta=50$ as the case in which all these effects
are maximised. For the same region of parameter space as before, 
the SM+2HDM is now around $-1.5.10^{-6}$ for $\mu>0$,
and a number between $-1.0$ and $-1.7.10^{-6}$ for $\mu<0$ (these
variations resulting from the significant increase in the contribution
of both the light neutral and the charged Higgs diagrams with respect 
to the $\tan \beta=2$ case);
in the presence of SUSY, we find that now the chargino contribution is
totally dominant, being well above both those of
the non-SUSY and the neutralino diagrams for both signs of $\mu$. This
can be seen in fig.~3(b)
which is analogous to fig.~3(a) but for $\tan \beta=50$.
As we can see, most of the $\mu<0$ plane and a considerable part of the
$\mu>0$ one give a total WMDM for the tau lepton an order of magnitude
well above the SM prediction (between $20$ and $6.10^{-6}$ for
$\mu>0$, and $-8$ and $-23.10^{-6}$ for $\mu<0$). They correspond to 
chargino contributions $a_{\chi^{\pm}} \sim 
(22 \rightarrow 7).10^{-6}$ ($\mu>0$) and $a_{\chi^{\pm}} \sim 
(-20 \rightarrow -6.5).10^{-6}$ ($\mu<0$), whereas the corresponding 
neutralino ones are almost negligible ($|a_{\chi^{0}}| \lesssim  
10^{-6}$).

\subsection{$(g-2)_{\mu}$}

It is interesting to relate these results for the WMDM of the $\tau$
lepton with the analogous calculation of $(g-2)_{\mu}$. This latter
case has been extensively studied in the literature, both for the SM
\cite{g2SM} and SUSY \cite{g2BSM} scenarios, and it is widely 
motivated by the increasingly good precision of recent and future 
measurements \cite{PDG,g2EXP}. As just said, the
calculation is totally analogous to that of the previous section, and
the formulae can be obtained from the appendix particularizing for
the case of having a photon and two muons in the external legs of the
different diagrams, which, among other things, allows an analytic 
solution of the Feynman integrals. We have computed the SUSY 
contribution to $(g-2)_{\mu}$ for exactly the same spectra for which
we computed the WMDMs of the previous section. Imposing a value 
$|\delta a_{\mu}| \leq 9.10^{-9}$ for the new contributions we find
that $\tan \beta \bigsim 10$ is the maximum allowed value for which these
spectra respect the bound on $\delta a_{\mu}$. Bigger values of $\tan
\beta$ will give rise to unacceptable values of the magnetic moment of
the muon.  
This can be clearly seen in fig.~4, where we plot $\delta a_{\mu}$
in the plane $\mu$ vs $M_2$ for (a) $\tan\beta=2$ and (b) $\tan \beta
=50$. 
\begin{figure}
\centerline{
\psfig{figure=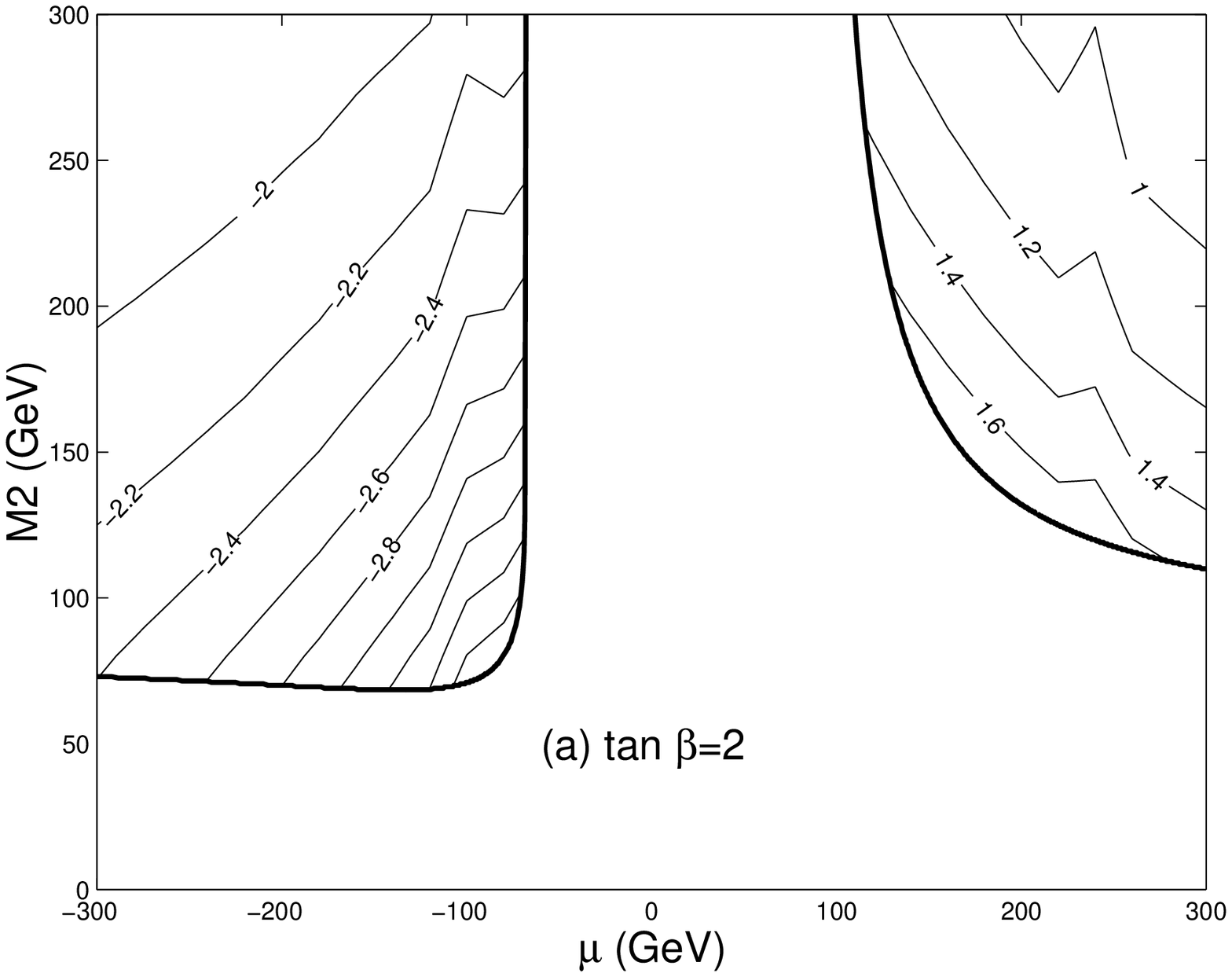,height=9cm,width=9cm,bbllx=0cm,bblly=7cm,bburx=21cm,bbury=21cm}\
\psfig{figure=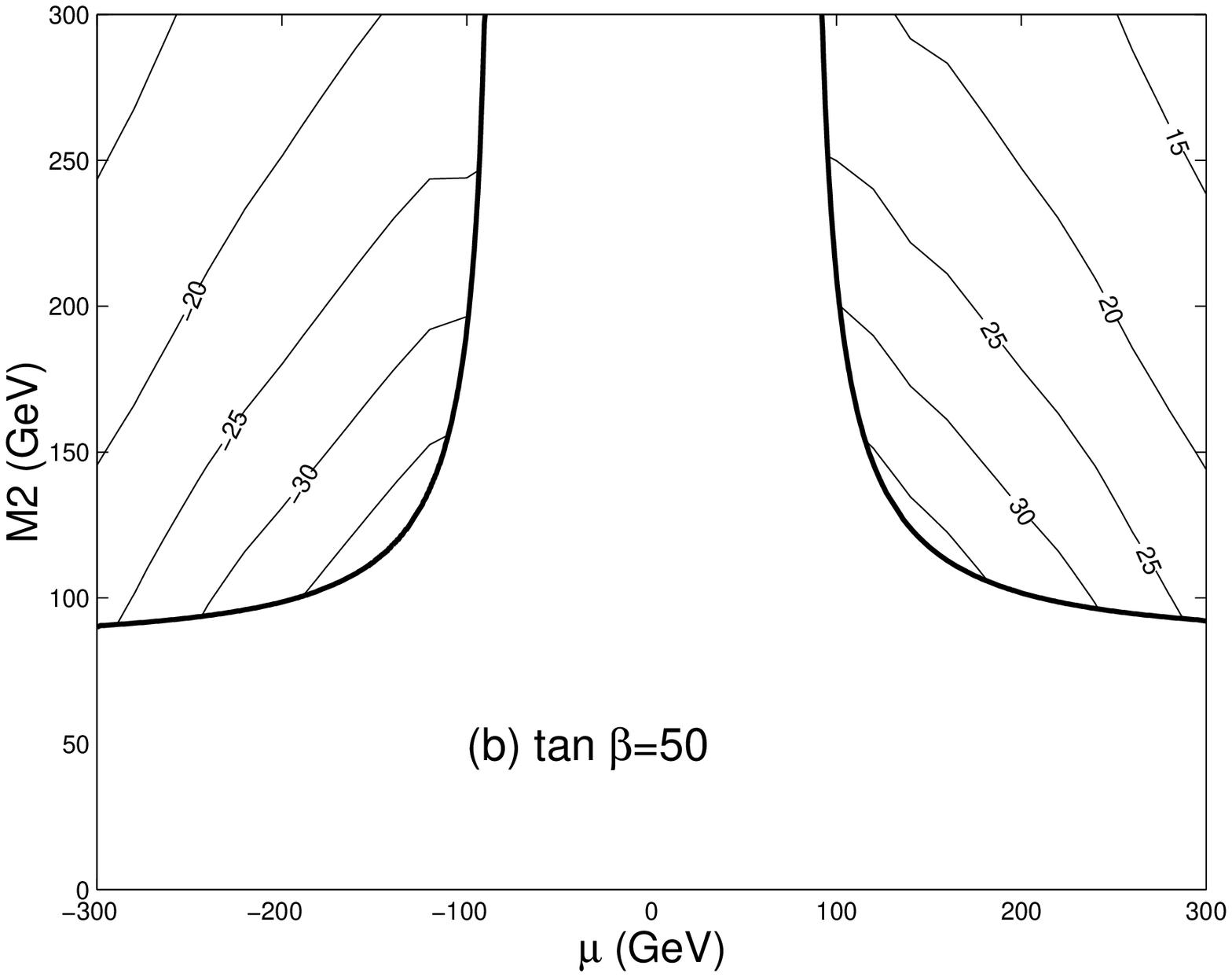,height=9cm,width=9cm,bbllx=0cm,bblly=7cm,bburx=21cm,bbury=21cm}
}
\caption{\footnotesize
Contour lines of constant $(g-2)$ (in units of $10^{-9}$) of the muon
 in the $\mu$ vs $M_2$ plane for (a) $\tan \beta=2$; (b) $\tan 
\beta=50$. The thick line in both plots represents the contour 
$m_{\chi^{\pm}_1} =85$ GeV. 
}
\end{figure} 
As for the WMDM of the $\tau$ lepton, the chargino/sneutrino
diagram is the dominant one, therefore it is the variation of 
$\delta a_{\mu}$ with those masses that is the most interesting 
one. As mentioned above, we can see that case (b) is totally ruled 
out by using the experimental bound also given above.

However, these results have been obtained under the assumption of
universal soft masses for the three families of squarks and sleptons,
that is, in our case $m_{L1}=m_{L2}=m_{L3}$, $m_{E1}=m_{E2}=m_{E3}$.
Essentially we are interested in keeping the outstanding predictions
for $a_{\tau}$ of the previous section for high values of $\tan \beta$
(over an order of magnitude above the SM result), while maintaning
$(g-2)_{\mu}$ under below its experimental bound. Given that in both
cases the dominant effect, as $\tan \beta$ increases, is due to the
chargino/sneutrino diagram, we may break the degeneracy between the
sneutrino masses associated with the different families, in order to 
achieve the desired result. Therefore we can compute how different
these masses have to be (or, in other words, how much bigger $m_{L2}$
has to be with respect to $m_{L3}$) to have a big WMDM for the $\tau$
lepton and an acceptable $(g-2)_{\mu}$. 

This can be easily read from fig.~5, where we plot both anomalous
magnetic moments vs a generic soft mass, $m_L$. The example
\begin{figure} 
\centerline{
\psfig{figure=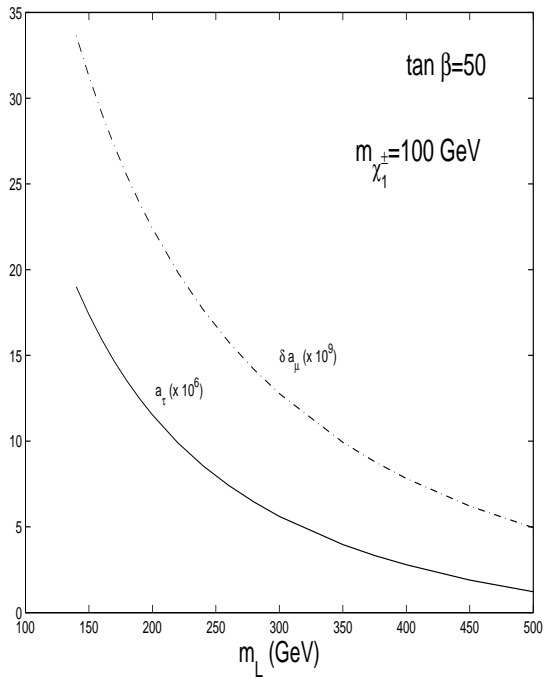,height=9cm,width=9cm,bbllx=0cm,bblly=7cm,bburx=21cm,bbury=21cm}
}
\caption{\footnotesize
Plot of the WMDM of the $\tau$ lepton (solid line, in units of 
$10^{-6}$) and of $(g-2)$ of the muon (dot--dashed line, in units of 
$10^{-9}$) vs a generic slepton soft mass, $m_L$, for $\tan \beta=50$. 
}
\end{figure} 
shown here corresponds to a lightest chargino mass of 100 GeV and
$\tan \beta=50$. To obtain a value of $\delta a_{\mu}$ below the
limit of $9.10^{-9}$ we have imposed (the most restrictive one)
the $m_{L2}$ must be above, approximately, 375 GeV. If we wish to 
preserve universality of the slepton masses, this in turn would give 
us a prediction for the WMDM of the $\tau$ lepton of $3.5.10^{-6}$.
However keeping $m_{L3}$ at some value between 100 and 200 GeV,
while driving $m_{L2}$ to the region of 400 GeV would result in
values of $a_{\tau} \sim (1.2-2.0).10^{-5}$, whereas the bounds
on $\delta a_{\mu}$ would be fulfilled. In general it will be 
possible to estimate to what extent this degeneracy has to be broken
in order to i) have $a_{\tau}$ as big as possible; ii) maintain
$|\delta a_{\mu}|$ below $9.10^{-9}$. The required $\Delta m_L \equiv
m_{L3}-m_{L2}$ will be a function of $\tan \beta$ and the value of
the lightest chargino mass we are picking to evaluate the relevant 
diagram in both cases.

The main direct effect of such lack of degeneracy of the slepton soft
masses would be the existence of non universal
corrections to the $Z$--lepton couplings \cite{Pilaftsis}.
These non-universal effects are strongly constrained by 
SLAC and LEP measurements (left-right asymmetries, leptonic widths, 
$\tau$ polarization data at the Z peak \cite{Toni}, etc.). First 
indications show that these corrections are never too large in SUSY 
extensions of the SM, but nevertheless the issue of universality 
breaking is interesting enough to deserve further investigation.

\section{The $b$ quark}

Let's now discuss very briefly the calculation of the WMDM for the
case of the $b$ quark. As already pointed out by the authors of 
ref.~\cite{BCLS96,BGV97},
the dominant contribution to this calculation comes from the gluon
diagram, which is two orders of magnitude bigger than the electroweak
ones. Altogether they give a total result of $(3.57-1.95i).10^{-4}$,
which will change very little when we plug in the SUSY 
contributions, which are given in the appendix. In general, the
inclusion of the two Higgs doublets will not have a remarkable impact
unless we are in a region of high $\tan \beta$ and low pseudoscalar
mass $m_A$ (for more details, see \cite{BCLS96}); therefore we will 
only discuss
the case of $\tan \beta=50$, which on the other hand, maximises the
SUSY contributions, which are now given by the diagrams analogous to the 
chargino and neutralino diagrams discussed for the $\tau$ lepton
(with stops and sbottoms, respectively, circulating in the loop) plus
a new diagram, a SUSY version of the dominant gluon one, which has
gluinos and sbottoms circulating in it.

\begin{figure} 
\centerline{
\psfig{figure=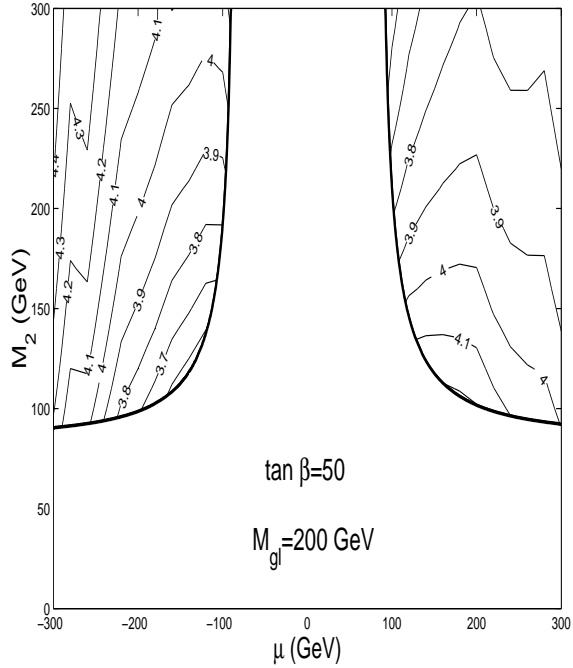,height=9cm,width=9cm,bbllx=0cm,bblly=7cm,bburx=21cm,bbury=21cm}
}
\caption{\footnotesize
Contour lines of constant WMDM of the $b$ quark (in units
of $10^{-4}$) in the $\mu$ vs $M_2$ plane, for $\tan \beta=50$ and 
$M_gl=200$ GeV. The thick line in the plot represents the contour 
$m_{\chi^{\pm}}=85$ GeV.
}
\end{figure} 
For $\mu<0$ $a_{non-SUSY}$ deviates from its pure EWSM+QCD value
due to the effect of the light Higgs diagram, which increases as the
absolute value of $\mu$ does. Therefore we have that $a_{non-SUSY}$
ranges from $3.7.10^{-4}$ to $4.3.10^{-4}$ when we go from $\mu=-100$
GeV to $\mu = -300$ GeV. If, on top of that, we include the effect 
of SUSY, as is shown in fig.~6, the total value does not change
much, because the effect of charginos and neutralinos, which
contribute with the same sign, is totally balanced by the gluino
diagram which has the opposite sign. $a_{gl}$ ranges between $39$ and 
$26.10^{-6}$ (for a representative gluino mass of 200 GeV), whereas 
$a_{\chi^0}$ lies between $-11$ and $-26.10^{-6}$, and the chargino 
contribution is the most variable one, given that it is the chargino 
mass which is changing on fig.~6: from $-48$ to $3.10^{-6}$.

As for $\mu>0$, $a_{non-SUSY}$ is almost constant at a value of
$3.9.10^{-4}$ (the light Higgs diagram is not sensitive to changes in
the value of $\mu$ as was also the case with the $\tau$ WMDM); the SUSY
diagrams now contribute with opposite signs to the $\mu<0$ case, but
there is no exact cancellation between chargino+neutralino diagrams
on one hand and the gluino diagram on the other. The final result is
then a total real part of the WMDM centered around $a_{non-SUSY}$,
being slightly smaller when the chargino contribution decreases (i.e.
big values of $M_2$ in fig.~6), and slightly bigger when the charginos
dominate. As we can read from fig.~6, $3.6.10^{-4} \leq a_{tot} \leq 
4.2.10^{-4}$. 

Finally, we have verified that, for the considered values of
the parameters, the SUSY corrections to $R_b$, also induced by these 
diagrams, are under control.

\section{Conclusion}

We have presented a full calculation of weak magnetic dipole moments
in the framework of the Minimal Supersymmetric Standard Model. In
particular we have performed a detailed analysis of the parameter
space for the case of the $\tau$ lepton and $b$ quark. For the former
we found that the SUSY contributions, in particular those of the   
diagrams involving charginos and sneutrinos, dominate over the SM  
result, especially for increasing $\tan \beta$. In general it is    
perfectly possible to achieve an enhancement of over an order of   
magnitude with respect to what the SM predicts. This in turn       
implies a too large contribution of the analogous SUSY diagrams    
to the calculation of $(g-2)$ of the muon when we assume universal   
slepton masses; however it is possible in principle to break this  
degeneracy without affecting any universality--breaking observables
and still keep this enhancement for the WMDM of the $\tau$
while having a prediction for $(g-2)_{\mu}$ within the experimental
limit.

\begin{flushleft}
{\large \bf Note added}
\end{flushleft}

After completion of this work we have received the paper of 
ref.~\cite{HIRS} where an independent analysis of the MSSM predictions  
for the WMDM of the ${\tau}$ lepton and $b$ quark is presented.

\section*{Acknowledgements}

BdeC thanks Luis Lavoura and Denis Comelli for very interesting 
discussions, and Mark Hindmarsh for his invaluable help in producing
most of the graphs. JMM thanks Alberto Casas for interesting 
discussions. Both BdeC and JMM thank, respectively, the 
Instituto de Estructura de la Materia (CSIC, Madrid) and the Centre 
for Theoretical Physics of the University of Sussex for hospitality 
during different stages of this work.

\appendixA{Appendix}

In this appendix we present the formulae for the supersymmetric 
contributions to the WMDM of a general, quark or lepton, fermion. 
As an example, we have depicted in fig.~2 the relevant diagrams 
involved in the calculation of the WMDM of the tau lepton.  A new 
type of diagram, associated with gluinos, has to be included if we 
calculate the WMDM of a quark. These diagrams can be expressed as a
combination of the scalar, vector and tensor Passarino-Veltman
three point functions \cite{PaV79}. We will use instead the 
parametrization given in \cite{BCLS96}, which we reproduce here
for the sake of completeness. Let $A,B,C$ be the tree particles
running into the loop, using the convention shown
in fig.~1. Then the relevant integrals are:

\br
I_{ 00 ;  \mu ;  \mu \nu } \; 
(p_-^2, (p_- + p_+)^2, p_+^2, m_A^2, m_B^2, m_C^2) & & =  \\
& & 
{ \displaystyle  \int   \frac {d^4 k} {i \pi} }
{ \displaystyle
 \frac {  \left \{ 1; k_{\mu}, k_{\mu \nu}  \right\} }
       { (k^2 -m_A^2) \left[(k-p_-)^2 - m_B^2 \right]
                      \left[(k+p_+)^2 - m_C^2 \right] }
}
\nonumber 
\er
which we will decompose as:
\be
\begin{array}{rcl}
I^{\mu}     &  =  &  (p_- - p_+)^\mu I_{10} + (p_- + p_+)^\mu I_{11} \\
I^{\mu \nu} &  =  &  (p_+^\mu p_+^\nu + p_-^\mu p_-^\nu)I_{21}  +
                     (p_+^\mu p_-^\nu + p_-^\mu p_+^\nu)I_{22}  + \\
&&                   (p_+^\mu p_+^\nu - p_-^\mu p_-^\nu)I_{2-1} +
                      g^{\mu\nu} I_{20}
\end{array}
\label{eq:Ies}
\ee
Let $a^{ABC}$ be the contribution to the WMDM given by this general 
diagram. Then  $a^{ABC}$  will be  expressed as a combination of the 
different $I_{ij} = I_{10}, I_{11}, I_{21}, I_{22} , I_{2-1}$
(see for example (A.4)). 
We omit some of the arguments of the $I_{ij}$s, 
since the external fermions are on shell.
To be more precise, we will use
\be
I_{ij} (m_A, m_B, m_C,q) \equiv 
I_{ij} (p_-^2, (p_- + p_+)^2, p_+^2, m_A^2, m_B^2, m_C^2) 
\ee
where $p_-^2=p_+^2 = m^2$, with $m$ the mass of the fermion we are 
considering. We have evaluated both the photon and Z-boson like 
magnetic dipole moments, $a^{\gamma}$ and $a^{Z}$, fixing $(p_-+p_+)^2 
\equiv q^2$ at $0$ and $M_Z^2$ respectively. Since the only difference, 
besides the $q$ value, is contained in the coupling of the gauge
boson, we will give a unique expression for the two cases.

               CHARGINO

Let us first consider the two diagrams associated to the charginos 
(see fig.~2). The contribution of the chargino-sfermion-sfermion 
diagram is given by:
%
%
%
\br
a^{\chi^- s s} & = &  - {\displaystyle \frac {\alpha_{em}}{4 \pi}}  
{\cal S}
 {\displaystyle \sum_{l=1,2}  \; \sum_{i,j=1,2} }  
 {\displaystyle \frac{m}{ s^2_W}}
( S_{j1}(x_s + y_s)S^*_{i1} + S_{j2}(x_s - y_s)S^*_{i2}) \nonumber
\\ & & \;\;\;\;\;\; \;\;\;\;\;\;
\left[ 2 m \, 
(\Lambda^{L*}_{jl} \Lambda^{L}_{il} + 
 \Lambda^{R*}_{jl} \Lambda^{R}_{il} ) \, (I_{10} - 
I_{21} + I_{22}) \right.   \label{eq:css}
\\
&  & \;\;\;\;\;\;\;\;\; \;\;\; \left.
+ M_{\chi^-_l} \,
(\Lambda^{R*}_{jl} \Lambda^{L}_{il} + 
 \Lambda^{L*}_{jl} \Lambda^{R}_{il} ) \, (I_{00} - 2 I_{10})
\right]
(M_{\chi^-_l},m_s^j,m_s^i,q)\, . \nonumber 
\er
%
The SUSY particles running in the loop are given in the mass basis, 
so the usual matrices relating mass and interaction states 
(for both charginos and sfermions) will appear in the vertices.
In particular,  $S_{i \alpha}$ is the corresponding rotation matrix
for the sfermions involved in the loop. The 
subscript $l$ labels the two charginos. The matrices $\Lambda^{R,L}$
mainly describe the chargino-sfermion-sfermion vertex. 
For stop-like squarks (i.e.,  $T_{3}^s = 1/2$), they are given by
\be
\begin{array}{rcl}
\Lambda^L_{il} & = & 
- {\displaystyle \frac{m_d} {\sqrt{2}M_W\cos\beta} } T_{i1} U_{l2}   
\\ [5mm]
\Lambda^R_{il} & = &  T_{i1} V^*_{l1}
- {\displaystyle \frac{m_u} {\sqrt{2}M_W\sin\beta} } T_{i2} V^*_{l2} \;\;,
\end{array}
\ee
with $S_{i \alpha}=T_{i \alpha}$, and for sbottom-like squarks 
($T_{3}^s = -1/2$)
\be
\begin{array}{rcl}
\Lambda^L_{il} & = & B_{i1} U^*_{l1}
- {\displaystyle \frac{m_d} {\sqrt{2}M_W\cos\beta} } B_{i2}  U^*_{l2}
\\ [5mm]
\Lambda^R_{il} & = &
- {\displaystyle \frac{m_u} {\sqrt{2}M_W\sin\beta} } B_{i1}  V_{l2} \;\;,
\end{array}
\ee
where $S_{i \alpha}=B_{i \alpha}$. 
The matrices $U,V$ in the previous equations are the usual
ones that allow expression of the two chargino mass eigenstates  
as a combination of the wino and charged higgsino. 
These expressions are trivially generalized for sleptons, by 
just replacing $m_u$ by $m_l$ and setting $m_d = 0$.

The coupling of the gauge boson $(Z, \gamma)$ to the sfermions 
is parametrized by
\be
x_s =
\left[
    \begin{array}{c}
        {\displaystyle \frac{1}{2s_W c_W} }
        (T_3^s - 2 Q_s s^2_W)
     \\ [3mm]
         Q_s
    \end{array}
\right]
\; \; \;
y_s =
\left[
    \begin{array}{c}
        {\displaystyle \frac{T_3^s}{2 s_W c_W} }
    \\ [3mm]
         0
   \end{array}
\right]
\ee

Finally, ${\cal S}$ stands for a global factor $2 T_{3f} = \pm 1$.

%
%
%

The contribution of the sfermion-chargino-chargino diagram is given by:
%
%
\br
a^{s \chi^- \chi^-} & = &  - {\displaystyle \frac {\alpha_{em}}{4 \pi}}     
{\displaystyle  \sum_{i=1,2} \;  \sum_{l,m=1,2} } 
{\displaystyle \frac{m}{s_W^2}}
\left[  
2 m
({\Lambda}^{L*}_{im} {\cal O}^{R}_{ml} {\Lambda}^{L}_{il}  +
 {\Lambda}^{R*}_{im} {\cal O}^{L}_{ml} {\Lambda}^{R}_{il} )
(I_{10} - I_{21}+I_{22})
\right. \nonumber  
\\ 
& & \;\;\;\;\;\;\;\;\;\; \left.   +
M_{\chi^-_l}
({\Lambda}^{R*}_{im} {\cal O}^{L}_{ml} {\Lambda}^{L}_{il} +
 {\Lambda}^{L*}_{im} {\cal O}^{R}_{ml} {\Lambda}^{R}_{il}  )
(I_{10} - I_{11}) \right. \label{eq:scc} \\ 
&  & \;\;\;\;\;\;\;\;\;\; \left.   +
M_{\chi^-_m}
({\Lambda}^{R*}_{im} {\cal O}^{R}_{ml} {\Lambda}^{L}_{il} +
 {\Lambda}^{L*}_{im} {\cal O}^{L}_{ml} {\Lambda}^{R}_{il}  ) 
(I_{10} + I_{11})
\right]
(m_s^i,M_{\chi^-_m},M_{\chi^-_l},q) \nonumber 
\er

where the matrices

\be
{\cal O}^L_{ml}  = 
\left[
    \begin{array}{c}
	{\displaystyle \frac{1}{s_W c_W}}
	(c^2_W \delta_{ml} - \frac{1}{2} V_{m2} V^*_{l2}) 
     \\ [3mm]
	 \delta_{ml}
    \end{array}  
\right]
\; \; \;
{\cal O}^R_{ml} = 
\left[
    \begin{array}{c}
	{\displaystyle \frac{1}{s_W c_W}} 
	(c^2_W \delta_{ml} - \frac{1}{2} U^*_{m2} U_{l2})  
    \\ [3mm]
	 \delta_{ml}
    \end{array}
\right]    
\ee
describe, up to some factor, the $(Z,\gamma)-\chi^+-\chi^-$
vertex.

		NEUTRALINO

The corrections induced by neutralinos are parametrized in a similar 
way. In particular, the contribution of the
neutralino-sfermion-sfermion diagram can be written as:
%
%
\begin{center}
\be
\begin{array}{rcrl}
a^{\chi^0 s s} & = &    {\displaystyle \frac {\alpha_{em}}{4 \pi}}
 {\displaystyle \sum_{l=1,4}  \; \sum_{i,j=1,2} }  &
  2 m \, ( S_{j1}(x_s + y_s)S^*_{i1} + S_{j2}(x_s - y_s)S^*_{i2})
\\ [4mm]
&&&
\left[ 2 m (
{\Lambda}_{jl}^{L*} {\Lambda}_{il}^{L} +
{\Lambda}_{jl}^{R*} {\Lambda}_{il}^{R} ) (I_{10} - I_{21}+I_{22}) 
\right.  
\\ [3mm]
&&&  
\left.   + 
M_{\chi^0_l} ( 
{\Lambda}_{jl}^{R*} {\Lambda}_{il}^{L} +
{\Lambda}_{jl}^{L*} {\Lambda}_{il}^{R} ) (I_{00} - 2 
I_{10}) \right]
(M_{\chi^0_l},m_s^j,m_s^i,q)\, . \label{eq:nss}
\end{array}
\ee
\end{center}
where now
\be
\begin{array}{rcl}
\Lambda^L_{il} & = &
\phantom{-}
   T_{i1} \left[
                  Q_u N'^*_{l1} +
                  {\displaystyle \frac{(v+a)_u}{2 s_W c_W} } N'^*_{l2}
          \right]
 + T_{i2} {\displaystyle \frac{m_u} {2 M_W s_W \sin\beta} }  N^*_{l4}
\\ [5mm]
\Lambda^R_{il} & = &
 - T_{i2} \left[
                  Q_u N'_{l1}   +
                  {\displaystyle \frac{(v-a)_u}{2 s_W c_W} } N'_{l2}
          \right]
 + T_{i1} {\displaystyle \frac{m_u} {2 M_W s_W \sin\beta} }  N_{l4}
\end{array}
\ee
for stop-like sfermions running in the loop, and
\be
\begin{array}{rcl}
\Lambda^L_{il} & = &
\phantom{-}
   B_{i1} \left[
                  Q_d N'^*_{l1} +
                  {\displaystyle \frac{(v+a)_d}{2 s_W c_W} } N'^*_{l2}
          \right]
 + B_{i2} {\displaystyle \frac{m_d} {2 M_W s_W \cos\beta} }  N^*_{l3}
\\ [5mm]
\Lambda^R_{il} & = &
 - B_{i2} \left[
                  Q_d N'_{l1}   +
                  {\displaystyle \frac{(v-a)_d}{2 s_W c_W} } N'_{l2}
          \right]
 + B_{i1} {\displaystyle \frac{m_d} {2 M_W s_W \cos\beta} }  N_{l3}
\end{array}
\ee 
for diagrams involving sbottom-like sfermions. In these equations,
$v = T_3 - 2 Q s^2_W$,  $a = T_3 $ and $N$ are the rotation matrices 
that relate mass eigenstate neutralinos to the photino, wino and 
neutral higgsinos.           

The sfermion-sfermion-neutralino diagram only contributes to $a^{Z}$
and is given by
%
%

\br
a^{s \chi^0 \chi^0} & = &  {\displaystyle \frac {\alpha_{em}}{4 \pi}}
{\displaystyle  \sum_{i=1,2} \;  \sum_{l,m=1,4} } 
{\displaystyle  \frac{2 m}{s_W c_W}}
\left[ 2 m
({\Lambda}^{L*}_{im} {\cal O}^{R}_{ml}{\Lambda}^{L}_{il}  +
 {\Lambda}^{R*}_{im} {\cal O}^{L}_{ml}{\Lambda}^{R}_{il} )
(I_{10} - I_{21}+I_{22})
\right. \nonumber \\  
& & \;\;\;\;\;\;\;\; \left.   +
M_{\chi^0_l}
( {\Lambda}^{R*}_{im} {\cal O}^{L}_{ml} {\Lambda}^{L}_{il} +
  {\Lambda}^{L*}_{im} {\cal O}^{R}_{ml} {\Lambda}^{R}_{il}  )
(I_{10} - I_{11})
\right.  \label{eq:snn} \\
& &  \;\;\;\;\;\;\;\; \left.   +
M_{\chi^0_m}
( {\Lambda}^{R*}_{im} {\cal O}^{R}_{ml} {\Lambda}^{L}_{il} +
  {\Lambda}^{L*}_{im} {\cal O}^{L}_{ml} {\Lambda}^{R}_{il}  )
(I_{10} + I_{11})
\right]
(m_s^i,M_{\chi^0_m},M_{\chi^0_l},q)\, . \nonumber
\er
where 
\be
\begin{array}{rcl}
{\cal O}^L_{ml} & = & 
- \frac{1}{2} N_{m3} N^*_{l3} + \frac{1}{2} N_{m4} N^*_{l4}  
\\ [3mm]
{\cal O}^R_{ml} & = & - {\cal O}^{L*}_{ml}
\end{array}
\ee
are the matrices describing the $Z-\chi^0-\chi^0$ vertex.
All these expressions are also valid for the WMDM 
of leptons, just making the replacements we suggested above.

                GLUINO
 
 Finally, we have to consider the gluino diagram in the evaluation of the
 WMDM of quarks. This diagram is similar to the 
 neutrino-sfermion-sfermion one. We get:

\br
a^{\tilde{g} s s} & = & {\displaystyle \frac {\alpha_s}{4 \pi}}
{\displaystyle \sum_{i,j=1,2} } 
C_R \, 2 m \,
( S_{j1}(x_s + y_s)S^*_{i1} +
  S_{j2}(x_s - y_s)S^*_{i2}) \label{eq:gss} \\ 
& & \;\;\;\;\;\;\;\; \left[
  2 m \, \delta_{ij} \,(I_{10}- I_{21} + I_{22}) -
M_{\tilde{g}} \, (S_{i2} S^*_{j1}
+ S_{i1} S^*_{j2}) \, (I_{00} - 2 I_{10}) \right]
(M_{\tilde{g}},m_s^j,m_s^i,q)\, . \nonumber
%
\er
%
where $C_R = 4/3$  and $ \alpha_s \equiv \alpha_s (M_Z) = 0.118$.

The previous expressions can be simplified
when considering $a^\gamma$. Firstly, 
some of the diagrams do not contribute.
Secondly, since the
$U(1)_{em}$ is unbroken, the photon coupling is of
course diagonal in the sfermion and chargino indices.
On the other hand, the $I_{ij}$ are reduced to
well known analytical functions when $q=0$ and
$m_B = m_C$. To be more precise, the following
relations hold:
%
%
\br
%
%
I_{11} (m,M,M,q)      & = & 0
\nonumber  \\ [5mm]
%
%
I_{10}  (m,M,M,0)                       & = &
-  {\displaystyle \frac{1}{4 M^2}}
\left [
(1 - 3x )(1-x)^{-2} -  2 x^2 (1-x)^{-3} \ln x
\right ] \nonumber
\\ [5mm]
%
%
[ I_{10}-I_{21}+I_{22} ] (m,M,M,0)      & = &
- {\displaystyle \frac{1}{12 M^2}}
\left [
(1 - 5x - 2 x^2)(1-x)^{-3} -  6x^2 (1-x)^{-4} \ln x
\right ] \nonumber
\\ [5mm]
%
%
[I_{00} - 2 I_{10} ] (m,M,M,0)          & = &
- {\displaystyle \frac{1}{2 M^2} }
\left [
(1 + x) (1-x)^{-2} +  2 x (1-x)^{-3} \ln x
\right ]
%
\er
%
where $x = (\frac{m}{M})^{2}$.
Then, the well known results for $(g-2)_\mu$ in SUSY models can be 
recovered as a particular and simple case in our general calculation.

\end{document}